\documentclass[aps,twocolumn,showpacs]{revtex4}%
\usepackage{amsfonts}
\usepackage{amsmath}
\usepackage{amssymb}
\usepackage{graphicx}%
\providecommand{\U}[1]{\protect\rule{.1in}{.1in}}
\providecommand{\U}[1]{\protect\rule{.1in}{.1in}}

\begin{document}
\title{Enhanced output entanglement with reservoir engineering}
\author{\ Xiao-Bo Yan}
\email{yxb@itp.ac.cn}
\affiliation{Institute of Theoretical Physics, Chinese Academy of Sciences, Beijing 100190, China}
\date{\today}


\begin{abstract}
We study the output entanglement in a three-mode
optomechanical system via reservoir engineering by shifting the center frequency of filter function away
from resonant frequency. We find the bandwidth of the filter function can
suppress the entanglement in the vicinity of resonant frequency of the system,
while the entanglement will become prosperous if the center frequency departs
from the resonant frequency. We obtain the approximate analytical expressions of the output
entanglement, and from which we give the optimal center frequency at which the
entanglement takes the maximum. Furthermore, we study the effects of time delay between the two output fields on the output entanglement, and obtain the optimal time delay for the case of
large filter bandwidth.

\end{abstract}
\maketitle

\section{Introduction}

Cavity optomechanics \cite{Aspelmeyer2014} exploring the interaction between
macroscopic mechanical resonators and light fields, has received increasing
attention for the potential to detect of tiny mass, force and displacement
\cite{Kippenberg2008,Marquardt2009,Verlot2010,Mahajan2013}. The common
optomechanical cavity contains one end mirror being a macroscopic mechanical
oscillator or a vibrating membrane
\cite{Gigan2006,Kleckner2006,Thompson2008,Groblacher2009,wenzhijia,xiaoboyan}. In these
optomechanical systems, the motion of mechanical oscillator can be effected by
the radiation pressure of cavity field, and this interaction can generate
various quantum phenomena. Such as ground-state cooling of mechanical modes
\cite{Marquardt2007,Wilson-Rae2007,Bhattacharya2007,Chan2011,Teufel2011,BingHe2017},
electromagnetically induced transparency and normal mode splitting
\cite{Huang2009,Weis2010,Safavi-Naeini2011,LiuYX2013,Kronwald2013}, nonlinear
interaction effects \cite{Komar2013,Lemonde2013,Borkje2013,LuXY2013} and
quantum state transfer between photons with vastly differing wavelengths
\cite{Tian2010,Stannigel2010,WangYD2012a,WangYD2012b}.

Entanglement is the characteristic element of quantum theory because it is
responsible for nonlocal correlations between observables and an essential
ingredient in most applications in quantum information. For these reasons,
there are a number of theoretical and experimental works on entanglement
between macroscopic objects such as, between atomic ensembles
\cite{Julsgaard2011,Krauter2011}, and between superconducting qubits
\cite{Berkley2003,Neeley2010,DiCarlo2010,Flurin2012}. Recently, quantum
entanglement in cavity optomechanics has received increasing attention for the
potential to use the interaction to generate various entanglement between
subsystems. For example, quantum entanglement between mechanical resonators
\cite{Bhattacharya2008,Chen2014,Liao2014,Yang2015}, between different optical
modes
\cite{Paternostro2007,Wipf2008,Genes2008,Barzanjeh2011,Barzanjeh2012,Barzanjeh2013,Wang2013,Tian2013,Kuzyk2013,Wang2015,Deng2015,Deng2016}%
, and between mechanical resonators and light modes
\cite{Vitali2007,Hofer2011,Akram2012,Sinha2015,BingHe1704} have been studied
theoretically and the entanglement between mechanical motion and microwave
fields has been demonstrated in a recent experiment \cite{Palomaki2013}.

Here, we consider a three-mode optomechanical system in which two cavities are
coupled to a common mechanical resonator (see Fig. 1). This setup has been
realized in several recent experiments \cite{Dong2012,Hill2012,Andrews2014}.
Because in such a system the parametric-amplifier interaction and the
beam-splitter interaction can entangle the two intracavity modes, the output
cavity ones are also entangle with each other. In previous works
\cite{Wang2015,Deng2016}, the entanglement of two output optical fields with
their center frequencies same as the resonant frequencies of the cavities has been studied. In Ref.
\cite{Wang2015}, the entanglement between the two output fields is enhanced
obviously via reservoir engineering \cite{Poyatos1996,Muschik2011}: cooling
the Bogoliubov mode through enhancing mechanical decay results in large
entanglement between the two target output fields. But these output
entanglement in Ref. \cite{Wang2015,Deng2016} will be largely limited by the
bandwidth of filter function, and the optimal time delay in Ref.
\cite{Wang2015} between the two output fields only suitable for the case of
little bandwidth of filter function.

In this paper, we first study the effect of filter bandwidth on the output entanglement between the two optical fields without time delay. We find the bandwidth will strongly suppress the output entanglement, specifically as the center frequency
of the output fields in the vicinity of resonant frequency.
While the output entanglement will become prosperous if the center frequency
of output fields departs from the resonant frequency. We will see that the physics behind this phenomenon is the reservoir
engineering mechanism because shifting the center frequency can cool the temperature of the system. We
obtain all the approximate analytical expressions of the output entanglement
in various case, and from which we give the corresponding optimal center
frequencies making the entanglement maximum. Finally, we study the effect of the time delay between the two output
fields on the output
entanglement according to the reservoir engineering mechanism, from which we obtain
the approximate analytical expression of the optimal time delay for the case
of large filter bandwidth. We think the results of this paper may be used for
reference to experimental and theoretical physicists who work on entanglement
or quantum information processing.

The rest of this paper is organized as follows. In Section II, we introduce
the three-mode optomechanical model with a corresponding equivalent model, and
the definition of canonical mode operators of the two output optical fields.
In Section III, we study the entanglement between the two output optical
fields by shifting the center frequency of filter function from resonant frequency. And we study the effects of time delay on the output entanglement. Finally, the conclusions are given in the Section IV.

\section{system and an equivalent model}

\begin{figure}[ptb]
\includegraphics[width=0.45\textwidth]{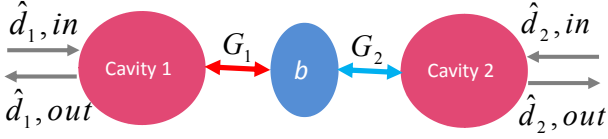}\caption{(Color online) A
three-mode optomechanical system with a mechanical resonator (mode $\hat{b}$)
interacted with two cavities (cavities 1 and 2). Cavity 1 is driven with a
red-detuned laser, while cavity 2 is driven with a blue-detuned laser. The
entanglement between the output fields of two cavities can be generated.}%
\label{Fig1}%
\end{figure}

We consider a three-mode optomechanical system in which two cavities are
coupled a common mechanical resonator (see Fig. 1).

The standard optomechanical Hamiltonian
\begin{align}
H  &  =\omega_{m}\hat{b}^{\dag}\hat{b}+\sum_{i=1,2}[\omega_{i}\hat{a}%
_{i}^{\dag}\hat{a}_{i}+g_{i}(\hat{b}^{\dag}+\hat{b})\hat{a}_{i}^{\dag}\hat
{a}_{i}] \label{Eq1}%
\end{align}
governs the system's dynamics, where $\hat{a}_{i}$ is the annihilation
operator for cavity $i$ with frequency $\omega_{i}$ and damping rate
$\kappa_{i}$, $\hat{b}$ is the annihilation operator for mechanics resonator
with frequency $\omega_{m}$ and damping rate $\gamma$, and $g_{i}$ is the
optomechanical coupling strength. In order to generate the steady entanglement
between the two output fields, we drive cavity 1 (2) at the red (blue)
sideband with respect to mechanical resonator: $\omega_{d1}=\omega_{1}%
-\omega_{m}$ and $\omega_{d2}=\omega_{2}+\omega_{m}$. If we work in a rotating
frame with respect to the free Hamiltonian, following the standard
linearization procedure, and make the rotating-wave approximation (in this
paper, we focus on the resolved-sideband regime $\omega_{m}\gg\kappa
_{1},\kappa_{2}$), hence, the Hamiltonian of the system can be written as
\begin{align}
\hat{H}_{int}  &  =G_{1}\hat{b}^{\dag}\hat{d}_{1}+G_{2}\hat{b}\hat{d}_{2}+H.c.
\end{align}
Here, $\hat{d}_{i}=\hat{a}_{i}-\bar{a}_{i}$, $\bar{a}_{i}$ being the classical
cavity amplitude. $G_{i}$ is the effective coupling strength. The combined
swapping and entangling interactions in $\hat{H}_{int}$ lead to a net
entangling interaction between the two intracavity modes as discussed in
\cite{Wang2013}.

Based on Eq. (2), the dynamics of the system is described by the following
quantum Langevin equations for relevant annihilation operators of mechanical
and optical modes%

\begin{align}
\frac{d}{dt}\hat{b}  &  =-\frac{\gamma}{2}\hat{b}-i(G_{1}\hat{d}_{1}+G_{2}%
\hat{d}_{2}^{\dag})-\sqrt{\gamma}\hat{b}^{in},\nonumber\\
\frac{d}{dt}\hat{d}_{1}  &  =-\frac{\kappa_{1}}{2}\hat{d}_{1}-iG_{1}\hat
{b}-\sqrt{\kappa_{1}}\hat{d}_{1}^{in},\label{Eq3}\\
\frac{d}{dt}\hat{d}_{2}^{\dag}  &  =-\frac{\kappa_{2}}{2}\hat{d}_{2}^{\dag
}+iG_{2}\hat{b}-\sqrt{\kappa_{2}}\hat{d}_{2}^{in,\dag},\nonumber
\end{align}
In Eq. (3), $\hat{b}^{in},\hat{d}_{i}^{in}$ are the input noise operators of
mechanical resonator and cavity $i (i=1,2)$, whose correlation functions are
$\langle\hat{b}^{in}(t)\hat{b}^{in,\dag}(t^{\prime})\rangle=N_{m}%
\delta(t-t^{\prime})$ and $\langle\hat{d}_{i}^{in}(t)\hat{d}_{i}^{in,\dag
}(t^{\prime})\rangle=N_{i}\delta(t-t^{\prime})$ respectively. Here, $N_{m}$
and $N_{i}$ are the average thermal populations of mechanical mode and cavity
$i$, respectively. In the following discussion, we mainly concentrate on how
the effects of the center frequency departing from the resonance, the
bandwidth of filter function on the entanglement, so we assume these average
thermal populations are zero (zero temperature). According to the
Routh-Hurwitz stability conditions \cite{DeJesus1987} and we focus on the
regime of strong cooperativities $C_{i}\equiv4G_{i}^{2}/(\gamma\kappa_{i}%
)\gg1$ and $\kappa_{i}\gg\gamma$ in this paper, the stability condition of our
system can be obtained as $G_{1}^{2}/G_{2}^{2}>\max(\kappa_{1}/\kappa
_{2},\kappa_{2}/\kappa_{1})$ for $\kappa_{1}\neq\kappa_{2}$, and the system is
always stable if $\kappa_{1}=\kappa_{2}$ and $G_{2}\leq G_{1}$
\cite{Wang2013,Wang2015}.

For simplicity, we adopt a rectangle filter with a bandwidth $\sigma$ centered
about the frequency $\omega$ to generate the output temporal modes. Then, the
canonical mode operators of the two output fields can be described as%

\begin{align}
\hat{D}_{i}^{out}[\omega,\sigma,\tau_{i}]=\frac{1}{\sqrt{\sigma}}\int%
_{\omega_{-}}^{\omega_{+}}d\omega^{\prime}e^{-i\omega^{\prime}\tau_{i}}\hat
{d}_{i}^{out}[\omega^{\prime}].
\end{align}
Here, $\omega_{\pm}=\omega\pm\frac{\sigma}{2}$, and $\tau_{i}$ is the absolute
time at which the wavepacket of interest is emitted from cavity $i$. The frequency-resolved output modes $\hat{d}_{i}^{out}%
[\omega]\equiv\int d\omega e^{i\omega t}\hat{d}_{i}^{out}[t]/\sqrt{2\pi}$
which can be obtained straightforwardly from the system Langevin equations and
input-output relations \cite{Gardiner2004}. And we use the logarithmic
negativity \cite{Vidal2002,Plenio2005} to quantify the entanglement between the
two output cavity modes $\hat{D}_{1}^{out}[\omega,\sigma,\tau_{1}]$ and $\hat{D}_{2}^{out}[-\omega,\sigma,\tau_{2}]$. Without
loss of generality, we set $\tau_{2}=0$, and we write $\hat{D}_{i}%
^{out}[\omega,\sigma,\tau_{i}]$ as $\hat{D}_{i}$ for simplicity in the
following.

It can be proofed that our system can be mapped to a two-mode squeezed thermal
state \cite{Wang2015}
\begin{align}
\hat{\rho}_{12}=\hat{S}_{12}(R_{12})[\hat{\rho}_{1}^{th}(\bar{n}_{1}%
)\otimes\hat{\rho}_{2}^{th}(\bar{n}_{2})]\hat{S}_{12}^{\dag}(R_{12})
\end{align}
Here,
\begin{align}
\hat{S}_{12}(R_{12})=\exp[R_{12}\hat{D}_{1}\hat{D}_{2}-H.c.]
\end{align}
is the two-mode squeeze operator, with $R_{12}$ being the squeezing parameter,
and $\rho_{i}^{th}(\bar{n}_{i})$ describes a single-mode thermal state with
average population $\bar{n}_{i}$. Hence, the output fields are thus completely
characterized by just three parameters: $\bar{n}_{1}$, $\bar{n}_{2}$, $R_{12}%
$. The relationship between the two-mode squeezed thermal state and our system
can be obtained as follows%

\begin{align}
\bar{n}_{1}  &  =\frac{\langle\hat{D}_{1}^{\dag}\hat{D}_{1}\rangle-\langle
\hat{D}_{2}^{\dag}\hat{D}_{2}\rangle-1+\sqrt{A^{2}-4|\langle\hat{D}_{1}\hat
{D}_{2}\rangle|^{2}}}{2},\nonumber\\
\bar{n}_{2}  &  =\frac{\langle\hat{D}_{2}^{\dag}\hat{D}_{2}\rangle-\langle
\hat{D}_{1}^{\dag}\hat{D}_{1}\rangle-1+\sqrt{A^{2}-4|\langle\hat{D}_{1}\hat
{D}_{2}\rangle|^{2}}}{2},\nonumber\\
R_{12}  &  =\frac{1}{2}\mathtt{arctanh}(\frac{2|\langle\hat{D}_{1}\hat{D}%
_{2}\rangle|}{A}),
\end{align}
here, $\langle\hat{D}_{1}^{\dag}\hat{D}_{1}\rangle$, $\langle\hat{D}_{2}%
^{\dag}\hat{D}_{2}\rangle$, $\langle\hat{D}_{1}\hat{D}_{2}\rangle$ are the
correlators of the output cavity modes, which can be obtained by Langevin
equations Eq. (3) and input-output relation, and $A=\langle\hat{D}_{1}^{\dag
}\hat{D}_{1}\rangle+\langle\hat{D}_{2}^{\dag}\hat{D}_{2}\rangle+1$. According
to Eq. (5) and Eq. (6), the output entanglement $E_{n}$ of this two-mode
squeezed thermal state (if $E_{n}\geq0$) can be simply given by
\begin{align}
E_{n}=-\ln(n_{R}-\sqrt{n_{R}^{2}-(1+2\bar{n}_{1})(1+2\bar{n}_{2})})
\end{align}
with $n_{R}=(\bar{n}_{1}+\bar{n}_{2}+1)\cosh2R_{12}$. It can be seen from Eq.
(8) that the entanglement will increase with the increase of the squeezing
parameter $R_{12}$, while decrease with the increase of the average
populations $\bar{n}_{1}, \bar{n}_{2}$. In the following, it can be seen that shifting the center frequency of filter function from the resonance can evidently cool the temperature of the system (decrease the average populations $\bar{n}_{1}, \bar{n}_{2}$).

\section{cavity output entanglement}

For simplicity, we set equal cavity damping rate $\kappa_{1}=\kappa_{2}%
=\kappa$, equal coupling $G_{1}=G_{2}=G$, and $\gamma\ll\sigma,\kappa,G$ in
the following. We discuss the output
entanglement on two cases: shifting the filter center frequency $\omega$ from the resonant
frequency (the resonant frequency is zero in the rotating frame) under the condition of small bandwidth ($\sigma\ll\kappa$), and large bandwidth ($\sigma=\kappa$) respectively.

\begin{figure}[ptb]
\includegraphics[width=0.45\textwidth]{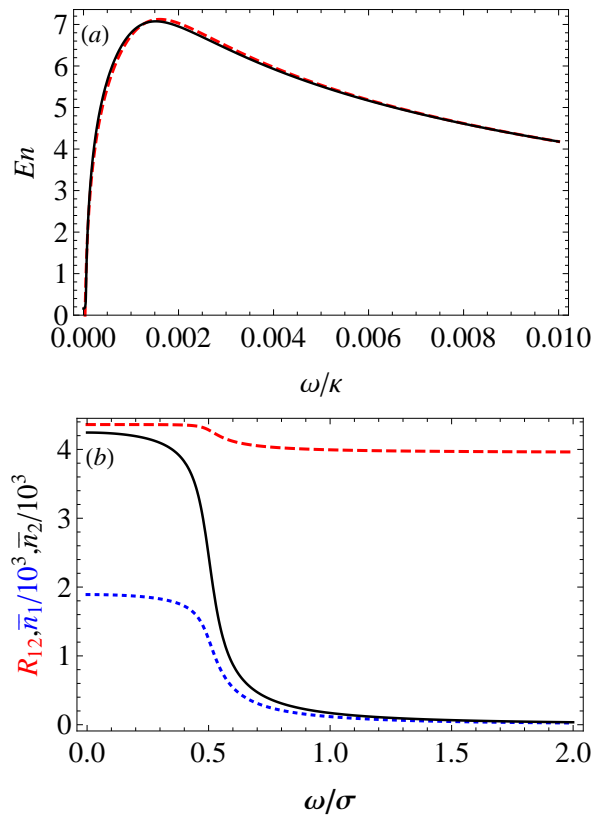}\caption{(a) The entanglement
vs the normalized center frequency $\omega/\kappa$. The black-solid line is
numerical result, the red-dashed line is plotted according to analytical
expression Eq. (10). (b) The squeezing parameter $R_{12}$ (red-dashed line),
the thermal populations $\bar{n}_{1}/10^{3}$ (blue-dotted line), $\bar{n}%
_{2}/10^{3}$ (black-solid line) vs the normalized center frequency
$\omega/\sigma$. The parameters are $\gamma=1, \sigma=10, \kappa=10^{5},
G=\kappa/10$.}%
\label{Fig2}%
\end{figure}

\subsection{Small bandwidth}

In this section we discuss the effects of small bandwidth $\sigma$ ($\sigma\ll\kappa$) on the
entanglement between the two output fields. If we shift the filter center frequency $\omega$
to satisfy $0\leq\omega\leq\frac{\sigma}{2}$ (in the rotating frame), the approximate analytical expression of
the output entanglement can be written as%

\begin{equation}
E_{n}\approx\frac{\pi\gamma}{2\sigma}.
\end{equation}
It can be seen from Eq. (9) that the entanglement between output fields is not
related to the filter center frequency $\omega$ and the coupling strength $G$. And increasing the mechanical decay
rate $\gamma$ can enhance the output entanglement in the vicinity of resonant
frequency $\omega=0$ just as what the author did in Ref. \cite{Wang2015},
which is the reservoir engineering mechanism because increasing mechanical
decay rate $\gamma$ can cool the Bogoliubov mode \cite{Wang2015}. If the
mechanical damping rate $\gamma$ satisfies $\gamma\ll\sigma$, the entanglement will
almost equal to zero. It can also be seen from Eq. (9) that the output entanglement can be largely suppressed by increasing the filter bandwidth $\sigma$.

If the center frequency $\omega$ satisfies $\frac{\sigma}{2}<\omega
<\frac{\kappa}{2}$, and the coupling strength $G$ is weak coupling ($G<\kappa$),
the analytical expression of the entanglement can be simplified to%

\begin{equation}
E_{n}\approx-\ln\frac{20G^{4}\sigma^{2}+3\kappa^{2}\omega^{4}}%
{3\omega^{2}(64G^{4}+\sqrt{2}\kappa^{2}\omega^{2})}.
\end{equation}

\begin{figure}[ptb]
\includegraphics[width=0.45\textwidth]{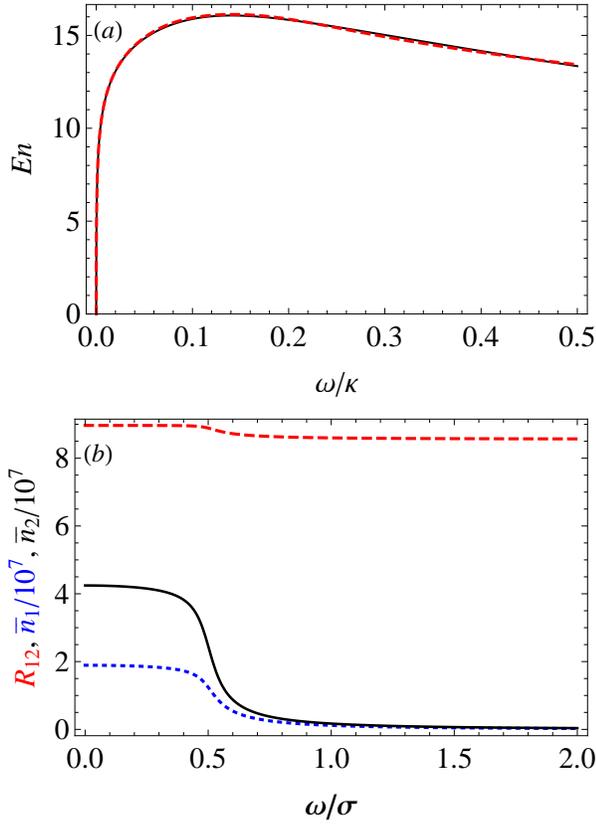}\caption{(a) The entanglement
vs the normalized center frequency $\omega/\kappa$. The black-solid line is
numerical result, the red-dashed line is plotted according to analytical
expression Eq. (11). (b) The squeezing parameter $R_{12}$ (red-dashed line),
the thermal populations $\bar{n}_{1}/10^{7}$ (blue-dotted line), $\bar{n}%
_{2}/10^{7}$ (black-solid line) vs the normalized center frequency
$\omega/\sigma$. The parameters are $\gamma=1, \sigma=10, \kappa=10^{5},
G=10\kappa$.}%
\label{Fig3}%
\end{figure}

The entanglement is plotted in Fig. 2(a) with parameters $\gamma=1, \sigma=10,
\kappa=10^{5}, G=\kappa/10$. The black-solid line is numerical result according
to logarithmic negativity, while the red-dashed line is plotted according to
simplified analytical expression Eq. (10). The entanglement is not monotonic
with the change of center frequency $\omega$, and will reach a maximum as the
optimal center frequency satisfy $\omega_{opt}\approx6^{1/4}G(\sigma
/\kappa)^{1/2}$. The entanglement will appear a peak value at resonant
frequency ($\omega=0$) for the case $\sigma=0$ \cite{Wang2015}, but the peak
will emerge at some a center frequency $\omega$ for the case $\sigma\neq0$.
We can give a clear reason for this phenomenon from Fig. 2(b) in which the
squeezing parameter $R_{12}$ (red-dashed line), the thermal populations
$\bar{n}_{1}$ (blue-dotted line), $\bar{n}_{2}$ (black-solid line) vs the
normalized center frequency $\omega/\sigma$ are plotted. It can be seen from
Fig. 2(b) the two thermal populations $\bar{n}_{1}$, $\bar{n}_{2}$ are very
large (the temperature of the equivalent two-mode squeezing thermal state is
very high) for $\omega<\sigma/2$, then the entanglement is almost zero. But if
the center frequency $\omega$ become larger ($\omega>\sigma/2$), the two
thermal populations $\bar{n}_{1}$, $\bar{n}_{2}$ will decrease rapidly while
the squeezing parameter $R_{12}$ decrease very slowly. Hence, the entanglement
become larger with the increase of center frequency $\omega$ until the highest
point. As a result, the optimal center frequency $\omega_{opt}$ at which the
entanglement reaches a maximum must be greater than $\sigma/2$.

If the coupling strength $G$ is strong coupling ($G>\kappa$), and the filter center frequency $\omega$ still satisfies $\frac{\sigma}{2}<\omega<\frac{\kappa}{2}$,
the analytical expression of the entanglement can be simplified to%

\begin{equation}
E_{n}\approx-\frac{1}{2}\ln[\frac{G^{8}\sigma^{4}+G^{4}\sigma^{2}%
\omega^{4}\kappa^{2}+2\omega^{10}\kappa^{2}}{144G^{8}\omega^{4}}],
\end{equation}
which reaches a maximum as the optimal center frequency satisfy
$\omega_{opt}\approx(G^{8}\sigma^{4}/3\kappa^{2})^{1/5}$. The entanglement is
plotted in Fig. 3(a) with parameters $\gamma=1, \sigma=10, \kappa=10^{5},
G=10\kappa$. The black-solid line is numerical result according to logarithmic negativity, while the red-dashed
line is plotted according to simplified analytical expression Eq. (11). It can
be seen from Fig. 2, Fig. 3 that the curves of entanglement plotted by
simplified analytical expressions fits the numerical results very well, the
squeezing parameter $R_{12}$ of strong coupling is larger than the case of
weak coupling, and the two thermal populations $\bar{n}_{1}$, $\bar{n}_{2}$ of
strong coupling will also decrease rapidly as the center frequency
$\omega>\sigma/2$ just as the case of weak coupling. That is the reason why
the entanglement of strong coupling will be larger than the one of weak coupling.

According to the above analysis that the optimal center frequency
$\omega_{opt}$ must be greater than $\sigma/2$, hence $\omega_{opt}$ will be
far away from the resonant frequency $\omega$ ($\omega=0$) if $\sigma$ is very large. We will discuss the case $\sigma=\kappa$ in the following.

\subsection{Large bandwidth}

\begin{figure}[ptbh]
\includegraphics[width=0.45\textwidth]{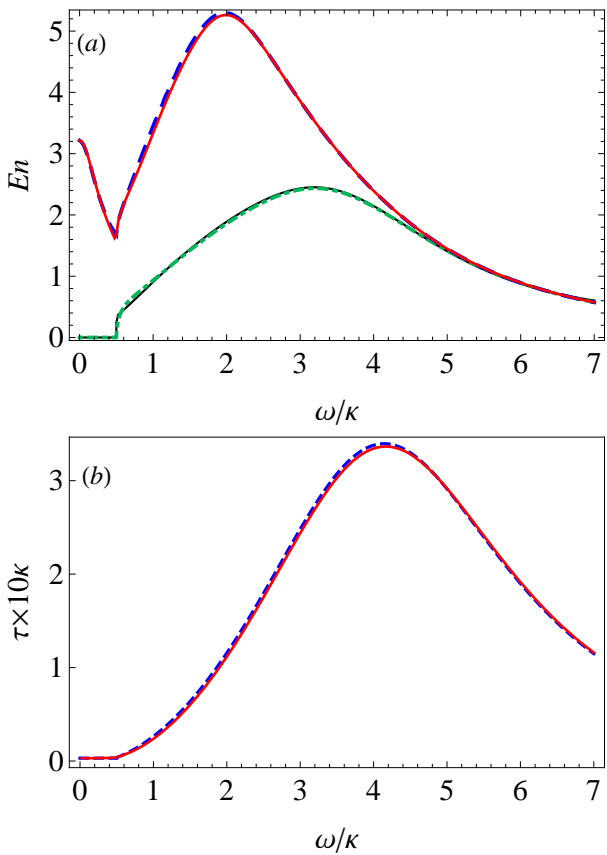}\caption{(a) The entanglement $En$
vs the normalized center frequency $\omega/\kappa$: The red-solid line is the entanglement plotted with the optimal time delay Eq. (14), the blue-dashed line is the entanglement plotted with the numerical optimal time delay making the entanglement $En$ maximum, the black-solid line is the entanglement plotted according to analytical expression Eq. (12) without time delay, and the green-dashed-dotted line is the entanglement plotted by numerical result according to the logarithmic negativity without time delay. (b) The optimal time delay $\tau_{opt}$ (red-solid line) according to Eq. (14) and the numerical optimal time delay (blue-dashed line). The parameters: $\gamma=1,\sigma=\kappa=10^{5},G=10\kappa$.}%
\label{Fig4}%
\end{figure}

For $G<\kappa$ and large $\sigma$, such as $G=\kappa/10$ and $\sigma=\kappa$,
the entanglement will be very small. Hence, in this section, we just discuss
the entanglement of strong coupling $G>\kappa$ with the bandwidth
$\sigma=\kappa$. Because of $\sigma=\kappa\gg\gamma$, the entanglement almost
be zero when $0\leq\omega\leq\frac{\kappa}{2}$ according to Eq. (9). The
analytical expression of the entanglement can be simplified to%

\begin{equation}
E_{n}\approx\ln[\sqrt{2}(\frac{3G^{4}\kappa^{2}(\omega^{2}+\frac
{3\kappa^{2}}{4})+G^{2}\kappa^{2}\omega^{4}+\omega^{8}} {3G^{4}\kappa
^{4}+2G^{2}\omega^{2}\kappa^{4}+\omega^{8}})]
\end{equation}
for $\frac{\kappa}{2}\lesssim\omega\lesssim7\kappa,$ and the optimal center
frequency $\omega_{opt}\approx\sqrt{G\kappa}$. In Fig. 4(a), we plot the
entanglement vs center frequency $\omega/\kappa$ according to the analytical
expression Eq. (9), Eq. (12) (black-solid line) and the numerical result
according to the logarithmic negativity (green-dashed-dotted line) under the
parameters: $\gamma=1,\sigma=\kappa=10^{5},G=10\kappa$. It can be seen from
Fig. 4(a) that there still is large entanglement even with large bandwidth ($\sigma=\kappa$). This because shifting center frequency can
effectively cool the two thermal populations $\bar{n}_{1}$, $\bar{n}_{2}$ via
reservoir engineering as above. And the tendencies of the two thermal populations
$\bar{n}_{1}$, $\bar{n}_{2}$ and the squeezing parameter $R_{12}$ are almost
the same as the previous cases in Fig. 2(b), Fig. 3(b), we don't discuss
them any more.

As the above analysis, large bandwidth $\sigma$ must strongly influence the entanglement of the two output fields. According to the definition of the canonical mode operators $\hat{D}_{i}$ (see Eq. (4)), the correlator of the output cavity modes $\langle\hat{D}_{1}\hat{D}_{2}\rangle$ is connected with time delay $\tau$, while the other two correlators $\langle\hat{D}^{\dag}_{1}\hat{D}_{1}\rangle$, $\langle\hat{D}^{\dag}_{2}\hat{D}_{2}\rangle$ are not. The expression $\langle\hat{D}_{1}\hat{D}_{2}\rangle$ can be written explicitly as

\begin{equation}
\langle\hat{D}_{1}\hat{D}_{2}\rangle=\int_{\omega_{-}}^{\omega_{+}}\frac{e^{-i\tau\Omega}(8G^{2}\kappa+(\gamma+2i\Omega)(\kappa
^{2}+4\Omega^{2}))}{-(\gamma^{2}+4\Omega^{2})(\kappa^{2}+4\Omega^{2})^{2}/(8G^{2}\kappa)%
}d\Omega.
\end{equation}
The effect of time delay $\tau$ on entanglement $E_{n}$ can be seen easily form the equivalent two-mode squeezing thermal state. From Eq. (7), we can see that the two-mode squeezing parameters $\bar{n}_{1}$, $\bar{n}_{2}$, and $R_{12}$ are affected by time delay $\tau$ just through the correlator $\langle\hat{D}_{1}\hat{D}_{2}\rangle$. More specifically, $\bar{n}_{1}$, $\bar{n}_{2}$ will decrease and $R_{12}$ will increase if the modulus $|\langle\hat{D}_{1}\hat{D}_{2}\rangle|$ becomes large as other parameters fixed except for time delay $\tau$. Hence, we can assert categorically that the output entanglement $E_{n}$ will increase with the increasing of the modulus of the
correlator $\langle\hat{D}_{1}\hat{D}_{2}\rangle$. The optimal time delay $\tau_{opt}$ is the delay which makes the $|\langle\hat{D}_{1}\hat{D}_{2}\rangle|$ reach a maximum. After obtaining the approximate analytical expression about $|\langle\hat{D}_{1}\hat{D}_{2}\rangle|$ and making some corrections, we find the optimal time delay is

\begin{equation}
\tau_{opt}\approx
\begin{cases}
\frac{3G^{2}\kappa(\omega^{2}-\frac{\kappa^{2}}{8})}{G^{4}\kappa^{2}%
+\omega^{6}}, & \omega\geq\frac{\kappa}{2}.\\
\frac{\pi\kappa}{2(2+\pi)G^{2}}, & 0\leq\omega<\frac{\kappa}{2}.
\end{cases}
\end{equation}

We plot the output entanglement $E_{n}$ with optimal time delay $\tau_{opt}$ (red-solid line) based on Eq. (14), and that with numerical optimal time delay which makes the entanglement $E_{n}$ reach a maximum (blue-dashed line) in Fig. 4(a) and the corresponding time delays are plotted in Fig. 4 (b) with the parameters: $\gamma=1,\sigma=\kappa=10^{5},G=10\kappa$, and they all fit very well. It can be seen from Fig. 4(a) that the time delay $\tau$ strongly affects the entanglement $E_{n}$ as long as the center frequency $\omega$ is not big enough compared with bandwidth $\sigma$, while has no effect on the entanglement $E_{n}$ as $\omega\gg\sigma$. The reason is that the effect of fixing $\sigma$ and increasing $\omega$ is equivalent to that of fixing $\omega$ and decreasing $\sigma$. And the time delay $\tau$ has no effect on entanglement for the case of $\sigma\rightarrow0$, which can be seen according to Eq. (13) that the factor $e^{-i\tau\Omega}$ can be extracted out of the integration for small bandwidth $\sigma$ with the result that the modulus $|\langle\hat{D}_{1}\hat{D}_{2}\rangle|$ will be not related to $\tau$. The steep entanglement in the vicinity $\omega=\sigma/2$ is because of the special rectangle filter and reaches a local minimum ($En_{min}\approx1.68$) at $\omega=\sigma/2$ according to the numerical result.

\section{Conclusions}

In summary, we have studied theoretically the output entanglement between two
output cavity fields via reservoir engineering by shifting the center
frequency of the causal filter function away from the resonance ($\omega=0$ in
the rotating frame) in a three-mode cavity optomechanical system. We find that the nonzero bandwidth $\sigma$ can largely suppress the
entanglement $En$, specifically in the vicinity of resonant frequency
$En\sim1/\sigma$. While the output entanglement will become prosperous, if we
shift the center frequency of output fields away from the resonant frequency.
This is because shifting center frequency can effectively cool the two-mode
squeezing thermal state which is equivalent to our model. We obtain all the approximate analytical expressions of the
output entanglement, and from which we give the corresponding optimal center
frequencies $\omega_{opt}$. In addition, we find the time delay $\tau$ between the two output optical fields can evidently effect the output entanglement. And we obtain the analytical
expression of the optimal time delay $\tau_{opt}$ in the case of large filter bandwidth ($\sigma=\kappa$). Our results can also be applied to other parametrically coupled three-mode bosonic systems, and may be useful to
experimentalists to obtain large entanglement.

\begin{acknowledgements}
XBY is supported by China
Postdoctoral Science Foundation (Grant No. 2015M571136).
\end{acknowledgements}

\end{document}